\documentclass{JHEP3} 
\usepackage{epsfig}
\usepackage{amsmath}

\usepackage{cite}

\newcommand{\beq}{\begin{eqnarray}}
\newcommand{\eeq}{\end{eqnarray}}
\newcommand{\bZ}[1]{\mathbb{Z}_{#1}}
\DeclareMathOperator{\tr}{tr} 
\newcommand{\EQ}[1]{\begin{equation} #1 \end{equation}}
\title {Confinement, Screening and the Center on ${\mathbf {S^3 \times S^1}}$}
\author{Carlos Hoyos${}^1$, Biagio Lucini${}^1$ and Asad Naqvi${}^{1,2}$\\
 1. Department of Physics, Swansea University, Singleton Park, Swansea SA2 8PP, UK.\\
 2.  Lahore University of Management and Sciences, Lahore, Pakistan. \\
        E-mail: \email{c.h.badayoz@swansea.ac.uk}, \email{b.lucini@swansea.ac.uk},\\ \phantom{E-mail:} \email{a.naqvi@swansea.ac.uk.}}
\abstract
{
We compute the one-loop effective potential for the Polyakov loop on  $S^3 \times S^1$ for an asymptotically free gauge theory of arbitrary group $G$ and a generic matter content. We apply this result to study the phase structures of $G_2$, $SO(N)$ and $Sp(2N)$ gauge theories which turn out to be in qualitative agreement with the results of lattice calculations. On $S^3 \times S^1$, at zero temperature, the Polyakov loop is zero for kinematical reasons. For small but non-zero temperature, the Polyakov loop is still zero
if the gauge theory has an unbroken center, while it acquires a small vacuum expectation
value for gauge theories whose center is trivial or explicitly broken by the presence of dynamical
matter fields. At high temperatures, 
the saddle point structure of the effective potential is different from the low temperature case suggesting that the theory is in the deconfined phase.  At finite $N$, the Polyakov loop is non-zero in the high temperature phase only in theories with no unbroken center symmetry, consistent with screening of the external charge introduced by the Polyakov loop and Gauss law on a compact space. 
}
\preprint{}
\begin{document}
\section{Introduction}
Quantum Chromodynamics (QCD) is a confining theory: the elementary
particles of the theory, quarks and gluons, never appear as final states
of strong interactions. This property holds for temperatures below a
critical value $T_c$, above which the system is deconfined, and quarks
and gluons form a plasma. Confinement is a property of both QCD and the
pure gauge theory without any matter.  To gain a better understanding of
this phenomenon, it is useful to study confinement in pure Yang-Mills
theory. For pure $SU(N)$ gauge theory, the existence of a deconfinement
phase transition has been rigorously proved in~\cite{Borgs:1983wb}.

The finite temperature theory is formulated on a four dimensional
Euclidean space with one compact dimension. The length $\beta$ of the
compact dimension is the inverse of the temperature $T$. In the confined
phase, the free energy of an isolated static quark is infinite. The free
energy $F$ is related to the vacuum expectation value of the Polyakov
loop
\beq
 L_R(\vec{x}) = \frac{1}{d_R}\, {\rm tr}_R\, \left( e^{i g \int_0^{1/T} A_0(\vec{x},t) d t} \right) \ ,
\eeq
where $A_0$ is the vector potential in the compact direction
(parameterized by the coordinate $t$), $g$ is the coupling of the
theory, $d_R$ is the dimension of the irreducible representation under
which the quark transforms\footnote{In QCD, this is the fundamental
representation; it is useful however to study a more general case, in
which the quark is in a generic representation of the gauge group.} and
$\vec{x}$ is the position of the quark. If we define the vacuum
expectation value ({\em vev}) of $L$ (we drop the representation index
from now on) as
\beq
\langle L \rangle = \frac{1}{V} \left\langle \int \, L(\vec{x}) d^3 x \right\rangle  \ ,
\eeq
the relationship between $L(\vec{x})$ and the free energy of the quark is given
by~\cite{McLerran:1981pb} 
\beq
F = - T \log \langle L \rangle \ .
\eeq
In $SU(N)$ gauge theories, the behavior of $\langle L \rangle$ across the
deconfinement phase transition can be explained in terms of a symmetry.
This is the symmetry of the system under the $\bZ{N}$ center of the
group, which leaves the Lagrangian unchanged. Under a transformation $z
\in \bZ{N}$, $\langle L \rangle \to z \langle L \rangle$. In the
deconfined phase, $F$ is finite, i.e.~$\langle L \rangle \ne 0$. This
implies that in the deconfined phase the center symmetry is broken.
Conversely, in the confined phase $F$ is infinite, $\langle L \rangle =
0$ and the center symmetry is not broken. This leads to the natural
interpretation that confinement reflects a change in the property of the
vacuum under the $\bZ{N}$ center symmetry, the {\em vev} of the Polyakov
loop being the order parameter for the transition. This observation lead
Svetitsky and Yaffe to the well-known conjecture~\cite{Svetitsky:1982gs}
that relates the deconfinement phase transition in a four dimensional
$SU(N)$ gauge theory to the order-disorder transition in three
dimensional $\bZ{N}$ spin systems: since the properties of the
deconfinement transition are determined by the underlying $\bZ{N}$
symmetry, the gauge and the corresponding spin system are in the same
universality class when both transitions are second order.

Since confinement is a non-perturbative phenomenon, a fully
non-perturbative approach such as putting the theory on a space-time
lattice is mandatory for quantitative studies. For $SU(N)$ gauge theory,
recently it has been shown in~\cite{Aharony:2003sx} that important
insights about the physics of color confinement and the phase structure
of the theory can be obtained by studying the system on a $S^3 \times
S^1$ manifold, where the radius $\beta$ of the $S^1$ is connected to the
temperature in the usual way ($\beta = 1/T$) and the radius $R$ of the
$S^3$ provides the IR cutoff at which the running of the gauge coupling
freezes.  When $1/R$ is much larger than the dynamical scale of the
theory, the coupling is small and perturbative calculations are
reliable. It is then possible to compute an effective potential for the
Polyakov loop at one loop order in perturbation theory. Evaluating the
partition function of the system for $\beta \ll R$ (high temperature
regime) and $\beta \gg R$ (low temperature regime) yields information
about the phase structure of the theory. In confining gauge theories,
for a finite number of degrees of freedom a cross-over between the
confined and deconfined phase is observed, with the two phases being
characterized by a different structure of the minima of the free energy
(or, using a more field theoretical terminology, the {\em effective
potential}). In order to verify the existence of a real phase transition
at some critical value of the temperature, an infinite number of degrees
of freedom is needed. In standard thermodynamical calculations, this is
achieved by taking the infinite volume limit of the system. This limit
is obviously not possible for $S^3 \times S^1$, since the calculations
rely on $R$ being small. Instead, the thermodynamic limit is achieved
here in another way: by sending to infinity the number of elementary
degrees of freedom of the theory (one example in $SU(N)$ gauge theory is
to take the limit of infinite number of colors $N$ with fixed $g^2
N$~\cite{tHooft:1973jz}). Thus, the existence of a phase transition can
be rigorously proved only in some large $N$ limit; for finite $N$ or for
exceptional groups, weak coupling calculations on a $S^3 \times S^1$
manifold can only discover the existence of two regimes (confined and
deconfined), but cannot establish whether they are separated by a
cross-over or a real phase transition in infinite volume.

Weak coupling calculations on $S^3 \times S^1$ have been used for
determining the phase structure of various supersymmetric and
non-supersymmetric gauge
theories~\cite{Aharony:2005bq,Barbon:2006us,Hollowood:2006xb,Unsal:2007fb,Unsal:2007vu},
treating each case independently. In this paper, we shall show that a
general expression can be derived from the representation theory of the
Lie Algebra of the gauge group that allows one to compute the one-loop
effective potential for a generic gauge theory containing bosons and
fermions in arbitrary representations. Using our result, it is
straightforward to work out the phase structure of non-Abelian gauge
theories for any gauge group and given matter content. We shall show
that our computation reproduces the previously known results and we will
use it to determine the phase structure of other gauge theories that can
help us to understand the physics of color confinement, like $G_2$ gauge
theory, with no center, and $SO(N)$ gauge theory, whose center is either
trivial ($N$ odd) or $\bZ{2}$ ($N$ even) \footnote{In this paper, we
will exclusively study $SO(N)$ groups and not their covering groups,
Spin(N).}. While a priori it is not obvious that the small volume
calculations reproduce features of the thermodynamical limit,
comparisons with other techniques (such as the lattice, as we will show
in this paper) provide evidence that this is indeed the case.

Our calculation show that in general the minimum structure of the effective potential is different at low and high temperatures. This means that a cross-over (in the finite volume case) or possibly a phase transition (which can arise only if the thermodynamical limit can be taken,  e.g. by sending to infinity the number of elementary degrees of freedom of the theory) separates those two regimes, which are naturally identified respectively with the confined and deconfined phase. However, in both regimes the details of the minimum structure depend on the center symmetry.
A symmetry cannot be spontaneously broken on finite volume, so if there
is a non-trivial center that is not broken explicitly by the matter
content of the theory, the Polyakov loop cannot develop an expectation
value at any temperature. Why the Polyakov loop vanishes in these theories differs in the
low temperature and the high temperature phase. In the low temperature
phase, the effective potential for the Polyakov loop will have only one
saddle point in which the Polyakov loop will be zero. At high
temperatures, there are multiple saddle points of the effective
potential, each related by the action of the center. In each of the
saddle points, the Polyakov loop is non-zero. However, since we are on
finite volume, these saddle points are not superselection sectors and
therefore we have to sum over all of them, which yields a zero value for
the Polyakov loop. It is plausible to assume that the existence of the
saddle points at high temperatures survives the infinite volume limit in
which, we can restrict ourselves to one of the multiple saddle points,
thereby breaking the center symmetry and signaling a
confinement/deconfinment transition. The situation is different when the
theory has no unbroken center subgroup. In that case, at both low and
high temperatures, there is one global minimum of the effective potential. In
the low temperature phase, the Polyakov loop is small $({\cal O}(e^{-\beta/R}))$ and is zero at zero temperature whereas in the high
temperature phase, we find that it develops an ${\cal O}(1)$ non-zero expectation
value. This can be understood in terms of screening the external charge
introduced by the Polyakov loop. For theories with a non-trivial center,
the external charge cannot be screened by gluons and quarks popping out
of the vacuum. However, if there is no center symmetry, any external
charge can be screened by an appropriate number of gluons and quarks.

The paper is organized as follows. In Sect.~\ref{generalities} we
compute the effective potential for a generic Lie group on a
$S^3 \times S^1$ manifold and derive the behavior of the Polyakov loop at
high and low temperature in terms of the root and weight
vectors of the group and the matter content of the theory, emphasizing how our
result reproduces previous calculations and how it can be used to analyze
other gauge theories. In Sect.~\ref{G2}, we specialize to the $G_2$ case,
showing that the Polyakov loop is zero at sufficiently low temperature and
different from zero at high temperatures. The same result is worked out
for $SO(N)$ gauge theories in Sect.~\ref{son}. In Sect. ~\ref{Gauss}   we show that our results are compatible with the non-Abelian Gauss law.
Finally, Sect.~\ref{conclusions}
sums up our findings and their implications for our understanding of
color confinement. 
\section{Confinement on  ${\mathbf{S^3}}$ for an arbitrary gauge theory}
\label{generalities}
In the canonical ensemble, the thermal partition function of a quantum
field 
theory is equal to the Euclidean path integral of the theory with a
periodic time direction of period $\beta = {1 \over T}$ and
anti-periodic boundary conditions for the fermions along the
temporal circle.  We will consider a gauge theory with a semi-simple
group $G$ with matter in arbitrary representations on a spatial $S^3$.  In the following, we will assume that the theories
we want to study are asymptotically free. To understand the phase
structure, we need to calculate the free energy which is just the log of
the partition function. Therefore, we perform a Euclidean path
integral for the theory on $S^3_R\times S^1_\beta$. 

We take $1/R$ to be large compared to the dynamical scale of the theory
, so the theory is at weak coupling. The $S^1$ topology allows
non-trivial flat connections $F_{\mu \nu}=0$ that can be parameterized
in a gauge invariant way by the {\em vev} of the Polyakov loop. The set
of flat connections are the lightest degrees of freedom of the theory,
and other excitations are separated in energy by a gap proportional to
$1/R$. In fact, the flat connection is a zero mode of the theory, in the
sense that the quadratic part of the action does not depend on it.  The low energy phase
structure can be studied using a Wilsonian approach where heavier
degrees of freedom are integrated out to give an effective action for
the flat connection. In the weak coupling regime, we can perform a
perturbative expansion in the coupling and the one loop result provides
a good approximation to the effective potential. Our discussion will
generalize previous results \cite{vanBaal:2000zc, Barbon:2005zj,
Barbon:2006us}. 

We can use time-dependent gauge transformations to fix $A_0$ to be constant and in the Cartan subalgebra,  
\begin{equation}
A_0=\sum_{a=1}^{\rm r} H^a C^a, \ \ [H^a,H^b]=0\   \forall a,b
\end{equation}
where $r$ is the rank of the group. The {\em vev} of the Polyakov loop is 
\begin{equation}
L=\,{\rm tr}\,\exp\left(\int_{S^1} A_0\right)\,.
\end{equation}
Notice that $A_0$ is anti-Hermitian in this expression. A coupling
constant factor has been absorbed in the field as well. This means that
in the classical Lagrangian the fields will not be canonically
normalized, but there is a $1/g^2$ overall factor. 

We will now consider adding matter fields to the gauge theory in
arbitrary representations that can be obtained as tensor products of the
fundamental (and anti-fundamental in case it exists). The fundamental
representation corresponds to a mapping of the group to the (complex)
general linear group acting on a vector space of dimension $F$.  So any
representation can be mapped to a tensor (we are ignoring the spinor
representations of $SO(N)$ groups). The tensor can have covariant or
contravariant indices, and the adjoint action can be defined as
(summation understood)
\begin{equation}
{\rm Ad}(A)\, X= A_{i_1}^{k_1} X_{k_1 i_2\cdots i_n}^{j_1 \cdots j_m}+\cdots+A_{i_n}^{k_n} X_{i_1 \cdots k_n}^{j_1 \cdots j_m}-A_{l_1}^{j_1} X_{i_1\cdots i_n}^{l_1 \cdots j_m}-\cdots -A_{l_m}^{j_1} X_{i_1\cdots i_n}^{l_1 \cdots l_m}\,.
\end{equation}
For $SU(N)$ groups, the Cartan subalgebra admits a diagonal form in
terms of the fundamental weights $\vec{\nu}_i$, $i=1,\dots,F$, where $F$
is the dimension of the fundamental representation.
\begin{equation}
H^a_{ij}=\nu_i^a \delta_{ij}\,.
\end{equation}
 The fundamental
weights need not be linearly independent; this will happen only if
$F=r$, the rank of the group. Real groups like $SO(N)$ and $Sp(N)$ cannot be
diagonalized using gauge transformations. However, once we have fixed
the flat connection to be constant, we can do a global similarity
transformation to diagonalize it. Gauge invariant operators remain
unchanged, so this transformation is allowed. In the case of $SO(N)$
this is done by a $U(N)$ transformation, while in the case of $Sp(N)$,
it is a $U(2N)$ transformation. In the diagonal basis, the adjoint
action simplifies to
\begin{equation}
{\rm Ad}(A)\, X=\vec{C} \cdot \left(\sum_{a=1}^n \vec{\nu}_{i_a} X_{i_1\cdots i_a \cdots i_n}^{j_1 \cdots j_m} - \sum_{b=1}^m \vec{\nu}_{j_b} X_{i_1\cdots i_n}^{j_1 \cdots j_b \cdots j_m}\right),
\end{equation}
which is 
\begin{equation}
{\rm Ad}(A)\, X = \vec{C} \cdot \left(\sum_{a=1}^n \vec{\nu}_{i_a} -\sum_{b=1}^m \vec{\nu}_{j_b} \right) X_{i_1 \cdots i_n}^{j_1\cdots j_m} = (\vec{C} \cdot \vec{\mu}_{ m,n} ) X_{i_1 \cdots i_n}^{j_1\cdots j_m}.
\end{equation}
Here, $\vec{C}$ , $\vec{\nu}$ and $\vec{\mu}_{m,n}$ are $r$ dimensional
vectors and the dot product is the standard Euclidean product. 
The weights of any representation are linear combinations of the
fundamental weights, so $ \vec{\mu}_{ m,n}= \left(\sum_{a=1}^n
\vec{\nu}_{i_a} -\sum_{b=1}^m \vec{\nu}_{j_b} \right) $ is a weight of
the tensor representation with $m$ covariant and $n$ contra-variant
indices.

The covariant derivative acting on a field in the background field of
the Polyakov loop is
\begin{equation}
D_\mu X = \partial_\mu X - \delta_{\mu 0}{\rm Ad}(A)\, X .
\end{equation}
To calculate the one-loop effective potential, we expand the gauge
field, matter and the ghosts in harmonics on $S^3 \times S^1$ and
keeping only the quadratic terms, we integrate out the various modes.
The one loop potential will be in general
\begin{equation}
V(C) = -{1\over 2} \sum_R (-1)^{F_R} {\rm Tr}_{\rm S^3\times S^1} \log \left(-D_{0R}^2- \Delta \right),
\end{equation}
where $R$ denotes the gauge group representation of the different fields
we have integrated out, $F_R$ is the fermionic number and the trace is
over the physical degrees of freedom (zero modes excluded). 
This can be written as  
\begin{equation}
V(C) = \sum_R (-1)^{F_R} \sum_{\vec{\mu}_R} V_R(\vec{\mu}_R\cdot \vec{C}),
\end{equation}
where $\vec{\mu}_R$ are the weights of the representation and the
potential $V_R$ can depend on the Lorentz representation of the field.
In flat space ($T^n\times R^d$ case) $V_R$ is just an universal function
times the number of physical polarizations (see \cite{Barbon:2006us},
for instance). On $S^3$ the situation is different because the
Kaluza-Klein decomposition depends on the spin of the field. The
one-loop potential for several $SU(N)$ theories on $S^3\times S^1$ has
been computed previously \cite{Aharony:2003sx,Hollowood:2006xb}. We will
generalize the computation to other groups and cast the results in a
simple form.

In general,
\begin{equation}
{\rm Tr}_{\rm S^3\times S^1} \log \left(-D_{0 R}^2-\Delta\right) = \sum_{\vec{\mu}_R} \sum_\ell \sum_n {\cal N}_R (\ell)\log\left( \varepsilon_\ell^2 + \left({2\pi n\over \beta}+ \vec{\mu}_R \cdot \vec{C}\right)^2 \right),
\end{equation}
where ${\cal N}_R$ is the number of physical polarizations of the field
and $\varepsilon_\ell$ is the energy of the Kaluza-Klein mode $\psi_l$
satisfying $\Delta \psi_l= -\varepsilon_\ell^2 \psi_l$ ($\Delta$ is the
Laplacian on $S^3$). ${\cal N}_R$ and $\varepsilon_\ell$ depend on the
field type that is being integrated out in a way that we list below. 

(i) {\bf Scalars}. Scalar field can be conformally or minimally coupled. The set of eigenvectors is the same in both cases. However, their eigenvalues are different. 
For conformally coupled 
scalars\footnote{These have a mass term involving the Ricci scalar of
  the manifold, in this case simply $R^{-1}$.} 
we have $\varepsilon_\ell=R^{-1}(\ell+1)$ whereas 
for minimally coupled scalars
$\varepsilon_\ell=R^{-1}\sqrt{\ell(\ell+2)}$. In both cases, there is a  
degeneracy ${\cal N}_R (\ell)=(\ell+1)^2$ with $\ell\geq0$. 

(ii) {\bf Spinors}. For 2-component complex
spinors, we have
$\varepsilon_\ell=R^{-1}(\ell+1/2)$ and
${\cal N}_R (\ell)=2\ell(\ell+1/2)$ with $\ell>0$. 

(iii) {\bf Vectors}.  
A vector field $V_i$ can be decomposed
into the image and the kernel of the covariant derivative:
$V_i=\nabla_i\chi+B_i$, with $\nabla^iB_i=0$. The eigenvectors 
for  $B_i$, have
$\varepsilon_\ell=R^{-1}(\ell+1)$ and ${\cal N}_R (\ell)=2\ell(\ell+2)$ with
$\ell>0$. 
On the
other hand, the  part  $\nabla_i\chi$ has
$\varepsilon_\ell=R^{-1}\sqrt{\ell(\ell+2)}$ with degeneracy
${\cal N}_R (\ell)=(\ell+1)^2$ but with $\ell>0$ only.  

Notice that both spinors and vectors have no zero ($\ell=0$) modes on
$S^3$. This is why in a pure gauge theory the only field with a zero
mode is $A_0$ which is a scalar on $S^3$.

It is a standard calculation using Poisson resummation
to show that up to an infinite additive constant, 
\EQ{
{\rm Tr}_{\rm S^3\times S^1} \log \left(-D_{0 R}^2-\Delta\right) = \sum_{\vec{\mu}_R}\sum_{\ell=0}^\infty {\cal N}_R (\ell)\Big\{\beta\varepsilon_\ell
-\sum_{n=1}^\infty\frac1n 
e^{-n\beta \varepsilon_\ell}\,\cos(n\beta ~ \vec{\mu}_R\cdot \vec{C})\Big\}\ .
\label{saa}
}
The first term here involves the Casimir energy and since it is independent
of $\vec{C}$ will play no role in our analysis and we will subsequently
drop it. Ignoring this term, the potential $V_R$ is given by

\begin{equation}
V_R(x) = 2\sum_\ell  {\cal N}_R (\ell) \sum_{n = 1} ^ \infty  e^{-n \beta \varepsilon_\ell } \;{\sin^2\left({n \beta \over 2}x\right)\over n}.
\end{equation}
The sum on $n$ can be done explicitly and the result is, up to a constant
\begin{equation}
V_R(x) = {1\over 2}\sum_\ell  {\cal N}_R (\ell) \log\left(1+{\sin^2\left({\beta x\over 2}\right)\over \sinh^2\left({\beta \varepsilon_\ell\over 2}\right)} \right)
\label{pot1}
\end{equation}
This is the contribution to the effective potential obtained by integrating out a particular field. The information about the type of field being integrated out (its Lorentz transformation properties etc) is  encoded in $  {\cal N}_R (\ell)$ and $\varepsilon_\ell$ as explained above. 

In the remainder of this section, we will restrict ourselves to a pure gauge theory with no matter fields. Then, the basic degree of freedom is the non-Abelian gauge potential $A_\mu$ which decomposes into a time component $A_0$ which transforms as a minimally coupled (massless) scalar on $S^3$  and a vector $A_i$ on $S^3$.  We can write $A_i = \nabla_i C + B_i$ such that $\nabla_i B^i=0$. Gauge fixing $ \Big(\nabla_iA^i+ 
D_0A_0 \Bigr)=0$ by the usual Fadeev-Popov procedure leads to the existence of ghosts.

Then, integrating out the $A_0$, $B_i$, $C$ and ghost modes can be performed using the following table:
$$
\begin{array}{ccccc}
{\rm field} & S^3\; {\rm KK-mode} & \varepsilon_\ell & ~~~~&{\cal N}_R (\ell)  \\
{ B_i} & \ell > 0 & (\ell+1)/R &~~~~ &2 \ell (\ell+2) \times(+1) \\
{ C} & \ell >0 & \sqrt{\ell(\ell+2)}/R & ~~~~&(\ell+1)^2 \times(+1) \\
{\rm ghosts} & \ell\geq 0 & \sqrt{\ell(\ell+2)}/R &~~~~ &(\ell+1)^2\times (-2) \\
{ A_0} & \ell\geq 0 & \sqrt{\ell(\ell+2)}/R &~~~~& (\ell+1)^2\times(+1) 
\end{array}
$$
Notice that for $\ell>0$, the modes $C$, ghost and $A_0$ cancel and the only contributions is from divergenceless modes, $B_i$. The constant $\ell=0$ modes for $C$, ghosts and $A_0$ give the following contribution:
\EQ{
V(x) \sim - \sum_{n = 1}^\infty {\cos(n   x)\over n }\sim \log\sin^2\left({ x\over 2}\right) 
}
This contribution is precisely the logarithm of the Jacobian that converts the integrals over $\vec{C}$ into the invariant integral over the group element $g = \exp \Bigl( i{\vec{\mu}_{\rm Ad} . \vec{C}} \Bigr)$, yielding the correct gauge invariant measure in the path integral $d\mu(g)$, $g\in G$. 
\begin{equation}\label{eq:haar}
{\cal Z}\simeq \int d\vec{C} \prod_{\mu_{\rm Ad}} \sin\left( {\vec{\mu}_{\rm Ad}\cdot \vec{ C}\over 2} \right) \cdots\simeq \int d\mu(g) \cdots \,.
\end{equation}
The contribution from the divergenceless part of $A_i$ which we have been denoting by $B_i$ can be easily written down using \eqref{pot1} and reading off the corresponding $\varepsilon_\ell$ and ${\cal N}_R (\ell) $ from the table above. 
 Then, the total effective potential is
\begin{equation}\label{eq:generalpotential}
V_{\rm gauge}(\vec{ C})=\sum_{\vec{\mu}_{\rm Ad}} {\cal V}(\vec{\mu}_{\rm Ad}\cdot \vec{ C})\,,
\end{equation}
where
\begin{equation}
{\cal V}(x)=-\log\left| \sin\left({x\over 2}\right)\right|+\sum_{l>0} l(l+2) \log\left(1+{\sin^2\left({x\over 2}\right)\over \sinh^2\left({\beta (l+1)\over 2 R}\right)} \right)\,.
\label{puregauge}
\end{equation}
The two terms in the effective potential come with opposite signs: the first term corresponds to a log repulsion coming from the $\ell=0$ modes whereas the second term is an attractive term. It is the interplay between these two types of contributions that leads to an interesting phase structure for the theory. 

In the low-temperature limit $\beta/R \gg 1$, the attractive contribution to the potential is exponentially supressed
\begin{equation}
{\cal V}(x)\simeq -\log\left| \sin\left({x\over 2}\right)\right|+\sum_{l>0} 4 l(l+2) \sin^2\left({x\over 2}\right)e^{-{\beta (l+1)\over  R}}+\dots\,.
\end{equation}
In the high-temperature limit $\beta/R \ll 1$, the lower mode terms become stronger and cancel the repulsive term except very close to the origin $x\sim \beta/R $
\begin{equation}
{\cal V}(x)\simeq -\log\left| \sin\left({x\over 2}\right)\right|+6 \log\left| \sin\left({x\over 2}\right)\right|+\dots \,.
\end{equation}
Therefore, we expect a change of behavior with increasing temperature. At low temperatures the expectation value of the Polyakov loop is zero up to exponentially supressed contributions. This can be seen from the fact that the repulsive term becomes the group measure (\ref{eq:haar}) and, since the Polyakov loop corresponds to the character of the group associated to the representation $R$,
\begin{equation}
\left\langle L \right\rangle = {1\over {\cal Z}} \int d\mu(g) \chi_R(g) =0\,.
\end{equation}
In the zero temperature limit this is a kinematical statement: there cannot be unscreened charges in a compact space. Therefore, confinement is just a kinematical fact. The derivation of the Haar measure in $S^3\times S^1$ was known for unitary groups \cite{Aharony:2003sx}, but our discussion is completely general and includes any semi-simple Lie group.

It is interesting to notice that at higher temperatures the expectation value of the Polyakov loop could be non-zero {\em even in finite volume}. An intuitive explanation follows from considering the Polyakov loop as the insertion of an external source in the system. At zero temperature pair production from the vacuum is completely frozen out, so it will cost an `infinite amount of energy' to introduce a charge in finite volume. However, at higher temperatures the system is in an excited state where the charges present in the medium can screen the external source. Whether this actually happens or not depends on the details of the group and the type of source we introduce. In gauge theories with unitary groups this issue boils down to whether the center of the group is broken or not. In pure Yang-Mills the center is unbroken, so the expectation value of the Polyakov loop should be zero at any temperature. On the other hand, in a theory with quarks the center is completely broken, so the Polyakov loop can be nonzero at high temperatures. In theories with an unbroken center,  the Polyakov loop has to be zero due to kinematics. However,  at high temperatures the main contributions to the path integral will come from a sum over saddle points corresponding to the minima of the potential. For very large temperatures the measure can be approximated by a sum of delta functions, so the value of the Polyakov loop is just the sum at the different minima. At each of these points the Polyakov loop has a nonzero value and it is plausible to assume that there is a map between them and the possible vacua of the gauge theory in the deconfined phase in infinite volume, where kinematical constraints do not apply. 
In this sense, by restricting ourselves to only one saddle point at high temperatures thereby getting a non-zero value of the Polyakov loop, we can interpret these results as a confinement-deconfinement phase transition at weak coupling. 

Strictly speaking there is no transition at finite volume, in the high temperature phase we should sum over all the vacua so the average of the Polyakov loop will be still zero or we would go smoothly from a zero to a non-zero value as we increase the temperature. However, it is possible to show there is a sharp transition in the large $N$ limit \cite{Aharony:2003sx}. Of course this is not applicable to the exceptional groups but only to the classical groups.

\subsection{Large $N$ phase transition}

In the ensuing analysis, it is convenient to write the potential in terms of the Polyakov loop eigenvalues $\lambda_i=\vec{\nu_i}\cdot \vec{C}$, where $\nu_i$ are the fundamental weights. This immediately follows from the fact that the weights of any tensor representation are linear combinations of the fundamental weights, and the contributions from integrating out such representations to the effective potential will depend on corresponding linear combinations of $\lambda_i$. In the $N\to \infty$ limit for the classical groups, we can approximate the discrete set of eigenvalues by a continuous variable $\theta$ and substitute the sums over eigenvalues by integrals in $\theta$ 
\begin{equation}
\sum_{\lambda_i} \to N \int d \theta \rho(\theta)\ ,
\end{equation}
with an eigenvalue density $\rho(\theta)\geq 0$, such that
\begin{equation}
\int_{-\pi}^\pi d \theta \rho(\theta) =1\ .
\end{equation}
The domain of integration is determined by the periodicity of the Polyakov loop eigenvalues. The potential becomes a functional of the eigenvalue distribution, and the system is well described by the saddle point that minimizes the potential. When the parameters of the theory change, such as the ratio $\beta/R$, the dominating saddle point can change, leading to a phase transition.

For pure $U(N)$ Yang-Mills, the potential is given by equation (\ref{puregauge}) with $x={\vec{\mu}_{\rm adj}}. \vec{C}= \lambda_i - \lambda_j$. After converting the summation to the integral via the density of eigenvalues, we obtain
\begin{equation}
V[\rho]=N^2\int_{-\pi}^\pi d \theta\rho(\theta) \int_{-\pi}^\pi d \theta' \rho(\theta') \sum_{k=1}^\infty {1\over k} (1-z_V(q^k)) \cos(k(\theta-\theta'))
\end{equation}
where $q=e^{\beta/R}$ and $z_V$ is the partition function of divergenceless vector fields in $S^3$
\begin{equation}
z_V(q)=\sum_{l>0} 2 l(l+2) q^{-(l+1)} = 2 {3\, q-1\over (q-1)^3}\ . 
\end{equation}
We can expand the eigenvalue distribution in complex Fourier components $\rho_n$
\begin{equation}
\rho(\theta)={1\over 2 \pi}+\sum_{n=1}^\infty {\rm Re}(\rho_n )\cos(n \theta)+\sum_{n=1}^\infty {\rm Im}(\rho_n )\sin(n \theta)
\label{fourier}
\end{equation}
and perform the integrals. We are left with a quadratic potential for each Fourier component
\begin{equation}
V=N^2 \pi^2 \sum_{n=1}^\infty {(1-z_V(q^n))\over n} |\rho_n|^2\ .
\end{equation}
For low enough temperatures,  $z_V(q^n) <1 $ for all $n$ and all the coefficients in the summation are positive so the configuration of minimal energy is the uniform distribution $\rho(\theta)=1/(2\pi)$. In this case the expectation value of the Polyakov loop vanishes, giving a weak coupling version of the confining phase. When we increase the temperature, the coefficient of the $n=1$ component first becomes zero at a critical value $q_c=2+\sqrt{3}$ and then negative beyond this point. If we continue increasing the temperature, eventually larger Fourier components become unstable beyond the points $q_c^{(n)}=q_c^{1/n}$ of marginal stability. At the critical value $q_c$ the eigenvalue distribution jumps to a qualitatively different saddle point characterized by the appearance of a gap in the eigenvalue distribution, {\em i.e.} there is a range of values on $\theta$ over which that function $\rho(\theta)$ has no support. At temperature just above $q_c$, 
\begin{equation}
\rho(\theta)={1\over 2 \pi}(1+\cos(\theta))\ .
\end{equation} 
The expectation value of the Polyakov in this saddle point is no longer zero, so this transition is a weak coupling analog of the confinement/deconfinement phase transition.

\subsection{Some examples}
We would like to emphasize that the formulas derived in this section are completely general on $S^3\times S^1$ for any semi-simple gauge group and for fields in any representation. We can cast the complete one-loop potential as a sum of different terms determined by the spin of the field and the weight system of the group representation. Possible examples are  pure gauge theories, with unitary, orthogonal symplectic or exceptional groups. We will comment some of them in the next sections.

Here we give a small list of examples in terms of the eigenvalues of the Polyakov loop $\lambda_i=\vec{\nu}_i\cdot \vec{C}$ for unitary groups. The phase diagram of most of them has been analyzed previously in the large-$N$ limit \cite{Aharony:2005bq,Unsal:2007fb}, showing an interesting structure in terms of the expectation values of the Polyakov loop. All of them exhibit a deconfinement transition at some critical temperature $T_c$. We will use the potential generated by Weyl fermions
\begin{equation}
{\cal V}_F(x)={1\over 2}\sum_{l>0}  l(2l+1) \log\left(1+{\sin^2\left({x\over 2}\,\left(+{\pi\over 2}\right)\right)\over \sinh^2\left({\beta (2l+1)\over 4 R}\right)} \right)\,.
\end{equation}
The $\pi/2$ addition in the argument is for anti-periodic boundary conditions on fermions. Then, we can write the following potentials:
\begin{itemize}
\item QCD: $SU(N)$ Yang-Mills with $N_f$ massless flavors (or theories with fundamental matter in general)
$$
 V(\lambda)= \sum_{i\neq j} {\cal V} \left(\lambda_i-\lambda_j \right) 
- 2 N_f\sum_{i=1}^N {\cal V}_F \left(\lambda_i\right)  \,.
$$
At high temperatures, the flavor contribution to the potential favors the $\lambda_i=0$ minimum over other center-breaking minima. At low temperatures the flavor potential is exponentially supressed, and only the repulsive term from the pure gauge contribution is relevant.

\item Supersymmetric theories: ${\cal N}=1$ $SU(N)$ (or theories with adjoint matter in general)
$$
 V(\lambda)= \sum_{i\neq j} {\cal V} \left(\lambda_i-\lambda_j \right) 
- \sum_{i\neq j} {\cal V}_F \left(\lambda_i-\lambda_j \right) \,.
$$
Notice that when $l\to \infty$, fermionic and bosonic contributions cancel at leading order if we choose periodic boundary conditions for fermions. At high temperatures all the center-breaking minima become deeper, so center symmetry is unbroken. At large $N$, the system will sit at one of these center-breaking minima. At low temperatures the fermionic contribution is exponentially supressed.

\item Orientifold theories \cite{Armoni:2003gp}: $SU(N)$ gauge theories with a Dirac fermion in the anti-symmetric (or symmetric) representation (or theories with two-index representations in general)
$$
 V(\lambda)= \sum_{i\neq j} {\cal V} \left(\lambda_i-\lambda_j \right) 
- \sum_{i\neq j} {\cal V}_F \left(\lambda_i+\lambda_j \right) \  -
\left(  2\sum_{i=1}^N {\cal V}_F \left(\lambda_i\right)  \right)\,.
$$
The term inside the  parentheses only appears in the symmetric representation; in the large-$N$ limit its contribution is sub-leading. At high temperatures there are two possible situations depending on $N$ being even or odd. In the first case, there are two-minima that become deeper w.r.t other center minima. If $N$ is odd, there is only one minimum. In general, the vacuum corresponds to non-trivial center vacua~\cite{Barbon:2005zj,Barbon:2006us}. At low temperatures we recover the pure gauge potential.

\item Orbifold theories (c.f.~\cite{Bershadsky:1998cb}), for instance the $\bZ{2}$ case with $SU(N)\times SU(N)$ gauge group and fermions in the bifundamental representation (or theories with several gauge group factors in general)
$$
 V(\lambda^{(1)},\lambda^{(2)})= \sum_{i\neq j} {\cal V} \left(\lambda_i^{(1)}-\lambda_j^{(1)} \right) +
\sum_{i\neq j} {\cal V} \left(\lambda_i^{(2)}-\lambda_j^{(2)} \right) 
- 2\sum_{i,j} {\cal V}_F \left(\lambda_i^{(1)}-\lambda_j^{(2)} \right) \,.
$$
In the limit $l\to \infty$ and large $N$, the diagonal sector $\lambda^{(1)}=\lambda^{(2)}$ becomes supersymmetric. At high temperatures the minima are on $\lambda_i^{(1)}=-\lambda_i^{(2)}$ \cite{Barbon:2005zj}, and in this sector, the potential is equivalent to the orientifold case.

\item Theories with more exotic matter content, etc. For a $SU(N)$ gauge theory, whenever we have a representation of non-trivial $N$-ality $\eta_R$, the center vacua will be shifted at high temperatures to a band with levels of $\eta_R$ degeneracy and a splitting proportional to $\eta_R/N$  \cite{Barbon:2006us}. At low temperatures any contribution is exponentially supressed unless conformally coupled scalars are included. In that case, there will be extra zero modes that can modify the low-temperature regime.

\end{itemize}
\section{The ${\mathbf{G_2}}$ case}
\label{G2}

The results of  section \ref{generalities} can be applied to a gauge theory with exceptional group $G_2$. We briefly summarize its relevant group theoretic properties (we refer to~\cite{Holland:2003jy} for a more in-depth discussion of $G_2$ gauge theory). $G_2$ has rank 2, so the potential (\ref{eq:generalpotential}) is defined over a two-dimensional plane. The fundamental representation is 7 and the adjoint 14 dimensional. The group $SU(3)$ is a subgroup of $G_2$.  The adjoint of $G_2$ under this embedding splits into an adjoint  and the fundamental and anti-fundamental representations of $SU(3)$: {\bf 14} $\to$ {\bf 8} $\oplus$ {\bf 3} $\oplus$ {\bf 3$^*$}. Therefore, 
the roots of $G_2$  are the same as the roots and the weights of the fundamental and anti-fundamental representation of $SU(3)$. 

The fundamental representation of $G_2$ can be decomposed in terms of $SU(3)$ representations as {\bf 7} $\to$ {\bf 1} $\oplus$ {\bf 3} $\oplus$ {\bf 3$^*$}. Then, the Polyakov loop of $G_2$ can be written in terms of a $SU(3)$ matrix $U$
\begin{equation}
L=\,{\rm tr}\, \left(\begin{array}{ccc} 1 & & \\ & U & \\ & & U^+ \end{array}\right)=1+ 2 \,{\rm Re} \,{\rm tr}\, U\,.
\end{equation}
As explained in section~\ref{generalities}, we choose to work in the gauge where the $SU(3)$ matrix for flat connections is parameterized by the constant Cartan subalgebra of the group
\begin{equation}
U=\,{\rm diag}\, \left(e^{i\vec{\nu}^1\cdot \vec{C}},\; e^{i\vec{\nu}^2\cdot \vec{C}},\; e^{i\vec{\nu}^3\cdot \vec{C}}\right)\,,
\end{equation}
 where the weights of the fundamental representation of $SU(3)$, $\vec{\nu}^i$, are not linearly independent but satisfy the following conditions:
\begin{equation}
\vec{\nu}^i\cdot \vec{\nu}^j=\delta^{ij}-{1\over 3}\ , \ \ \ \sum_{i=1}^3 \vec{\nu}^i=0\,.
\end{equation}
The physical region of the $\vec{C}$ plane is obtained quotienting by global periodic transformations, that shift the phase of the eigenvalues of the Polyakov loop by multiples of $2 \pi$:   $\vec{\nu}^i\cdot\vec{C}\to \vec{\nu}^i\cdot\vec{C}+2 \pi n^i$, and by Weyl transformations, that permute the eigenvalues of the Polyakov loop. In both cases, the root lattice determines the structure of the Cartan torus. Periodic transformations can be seen as translations along the roots $\alpha^{ij} = \nu^i-\nu^j$, while Weyl transformations are seen as the reflections that move fundamental weights among themselves. The fixed lines of Weyl reflections lie along the roots. As a basis for the Cartan torus we can choose the simple roots $\alpha_\pm=(1/\sqrt{2},\pm \sqrt{3/2})$. Since $\nu^i\cdot \alpha_\pm=0,1$, the period is $2\pi$. If we consider other representations, the group of Weyl reflections will be different and the associated Weyl lines will show a different structure. For a $SU(3)$ theory with fields in the adjoint representation we can also make $\bZ{3}$ center transformations that shift the eigenvalues by multiples of $2\pi/3$. On the Cartan torus, this is seen as translations along the weights by $2\pi$ units. In figures~\ref{fig:su3torus}, \ref{fig:su3pot} we show some examples.

\FIGURE{
\includegraphics[scale=0.45]{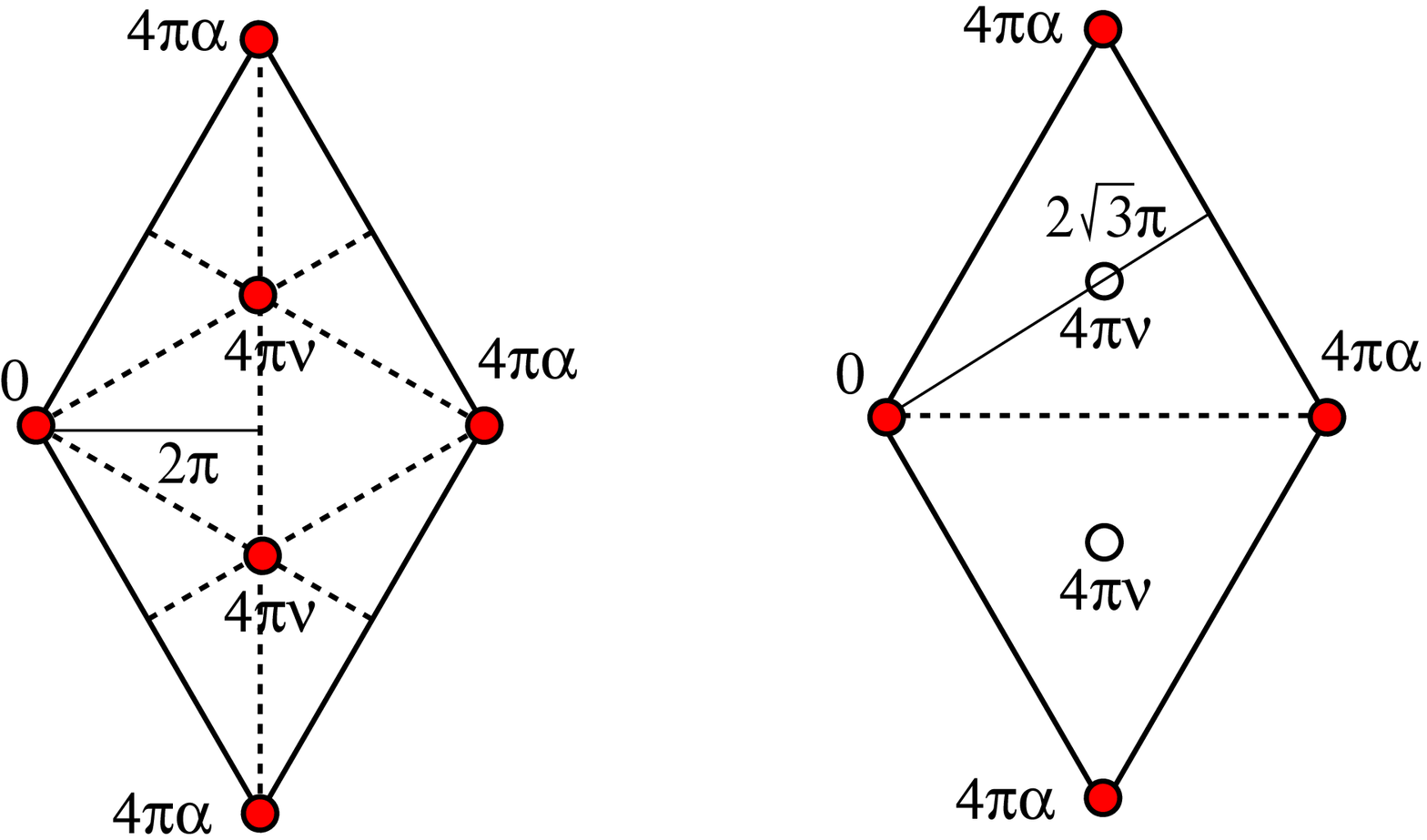}
\caption{\label{fig:su3torus}\small The Cartan torus (with roots normalized to unit norm) for $SU(3)$ for the adjoint representation (on the left) and the fundamental representation (on the right). The Weyl lines are indicated by dashed lines. The global minima of the potential lie on the intersection of Weyl lines. For the fundamental representation, we also mark by white circles the smooth critical points of the potential. The center symmetry corresponds to translations on the torus that move the global minima of the adjoint representation among each other, notice that it is not a symmetry of the fundamental.}
}
\FIGURE{
\begin{tabular}{cc}
\includegraphics[scale=0.45]{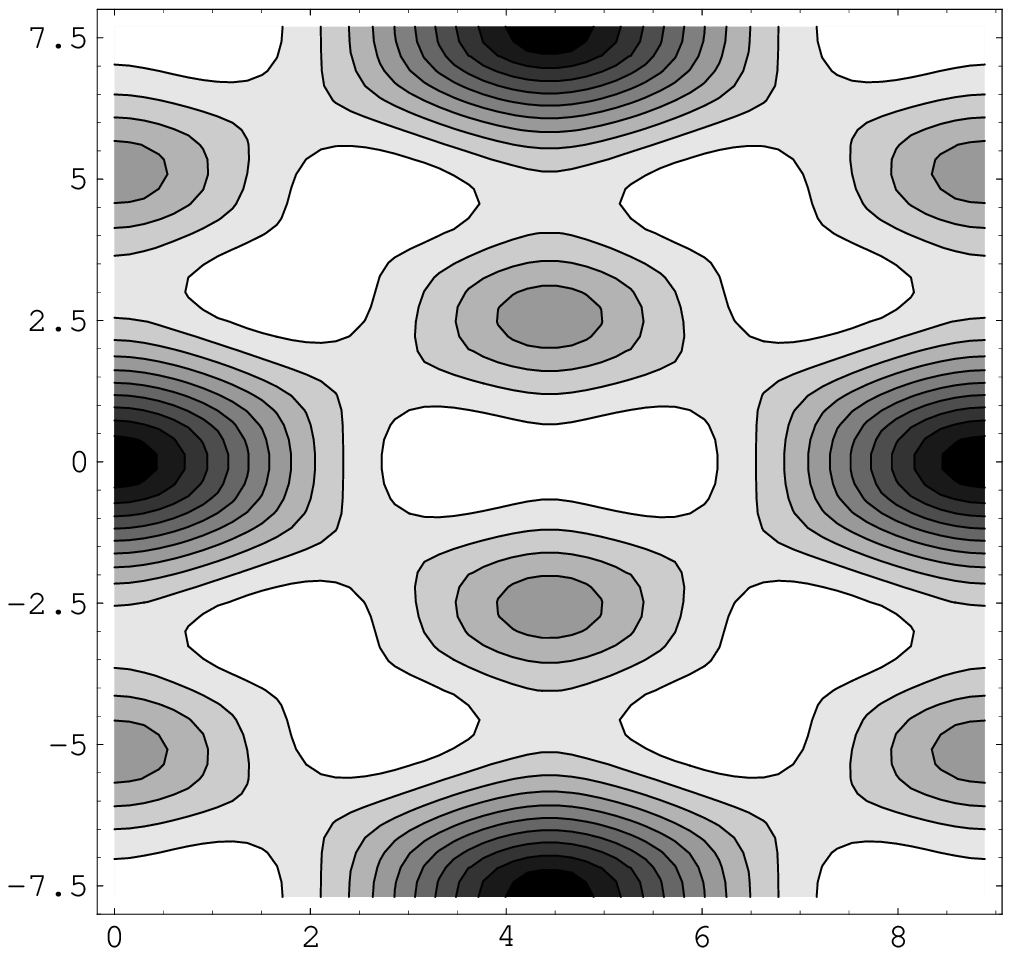} & \includegraphics[scale=0.55]{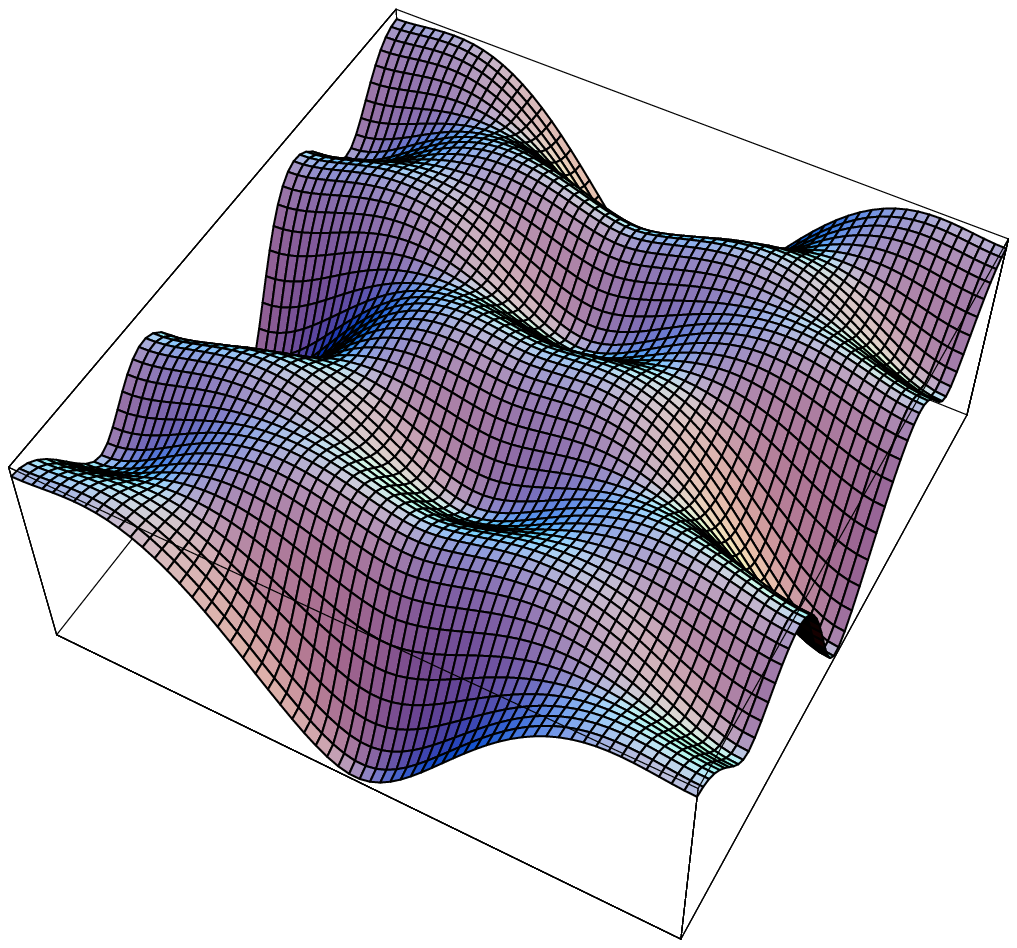} \\
(a) & (b)
\end{tabular}
\caption{\label{fig:su3pot}\small The potential for $G_2$ Yang-Mills at high temperatures. There is a global minima located at the origin of the Cartan torus and two local minima inside.}
}

The potential is then
\begin{equation}\label{eq:g2potential}
V(\vec{C})=2\sum_{i=1}^3 {\cal V}\left( \vec{\nu}^i\cdot\vec{C} \right)+\sum_{i\neq j} {\cal V}\left( (\vec{\nu}^i-\vec{\nu}^j)\cdot\vec{C}\right)\,.
\end{equation}
In the $SU(3)$ gauge theory,  the first term in (\ref{eq:g2potential}) does not appear. In that case, the attractive contribution of the potential has its minima located where the cosine is maximum, {\em i.e.} at
\begin{equation}
(\vec{\nu}^i-\vec{\nu}^j)\cdot\vec{C} = 2 \pi n\ , \ \ n\in \mathbb{Z}\ .
\end{equation}
For the adjoint representation, there are three minima located inside the physical region. In the pure Yang-Mills theory, one can move from one minimum to the other using $\bZ{3}$ center transformations. Therefore, the localization around the minima at high temperatures could be interpreted as the weak coupling analog of the breaking of the center in the infinite volume theory. When the temperature is very low, the repulsive contribution of the potential dominates, so the minima become maxima and we expect the configurations to be spread along the Cartan torus, giving a vanishing expectation value to the Polyakov loop. When we consider matter in the fundamental representation of $SU(3)$, or in the $G_2$ pure gauge theory, the center is no longer a symmetry. In the one-loop potential (\ref{eq:g2potential}), this is reflected by the presence of the first term. The minima of the attractive contribution are located at
\begin{equation}
\vec{\nu}^i\cdot\vec{C} = 2 \pi n\ , \ \ n\in \mathbb{Z}\ .
\end{equation} 
So there is only one minimum inside the physical region. Then, only one of the minima of the $SU(3)$ adjoint contribution is a global minimum, while the other two can only be local minima at most. However, the localization around the minimum still occurs when we change the temperature.

In figure~\ref{fig:g2potential} we have plotted the $G_2$ potential for different temperatures. We notice that in the high temperature phase there is a single global minimum located at the corners of the physical region. This minimum corresponds to the trivial vacuum, with ${\rm tr}\,U =1$. There are also two local minima that in the pure $SU(3)$ gauge theory will be the minima corresponding to the Polyakov loop values ${\rm tr}\, U= e^{\pm 2 \pi i/3}$.
When we lower the temperature, the minima become repulsive points due to the logarithmic term in the potential. The most repulsive point is the trivial vacuum, so the configurations will be distributed around the center values. 

We can compute the expectation value of the Polyakov loop at different temperatures. In terms of the eigenvalues $\lambda_i=\vec{\nu}_i\cdot \vec{C}$, the zero temperature $T=0$ path integral is
\begin{equation}
{\cal Z}_0= \left(\prod_{i=1}^3 \int_{-\pi}^\pi  d \lambda_i\right) \, \delta\left(\sum_{i=1}^3 \lambda_i \right) \prod_{i<j} \sin^2\left({\lambda_i-\lambda_j\over 2}\right) \prod_{i=1}^3 \sin^2\left({\lambda_i\over 2}\right)\equiv \int d\mu(\lambda) \,.
\end{equation}
The expectation value of the Polyakov loop is exactly zero
\begin{equation}
\left\langle L \right\rangle_{T=0} = {1\over {\cal Z}_0} \int d \mu(\lambda) \left(1 +2 \sum_{i=1}^3 \cos(\lambda_i)\right) =0\,.
\end{equation}
At non-zero temperature $T=1/\beta$, the Kaluza-Klein modes on $S^3$ give non-trivial contributions
\begin{equation}
{\cal Z}_\beta=\int d\mu(\lambda) \prod_{l=1}^\infty \left[\prod_{i<j} \left(1+{\sin^2\left({\lambda_i-\lambda_j\over 2}\right)\over \sinh^2\left({\beta (l+1)\over 2 R}\right)} \right) \prod_{i=1}^3 \left(1+{\sin^2\left({\lambda_i\over 2}\right)\over \sinh^2\left({\beta (l+1)\over 2 R}\right)} \right)\right]^{-2l(l+2)}\,. 
\end{equation}
The expectation value becomes nonzero, and for high temperatures the measure is peaked around $\lambda_i=0$. This qualitative picture agrees with recent lattice results \cite{Greensite:2006sm}.
\FIGURE{
\begin{tabular}{ccc}
\includegraphics[scale=0.45]{highTpot.eps} & \includegraphics[scale=0.45]{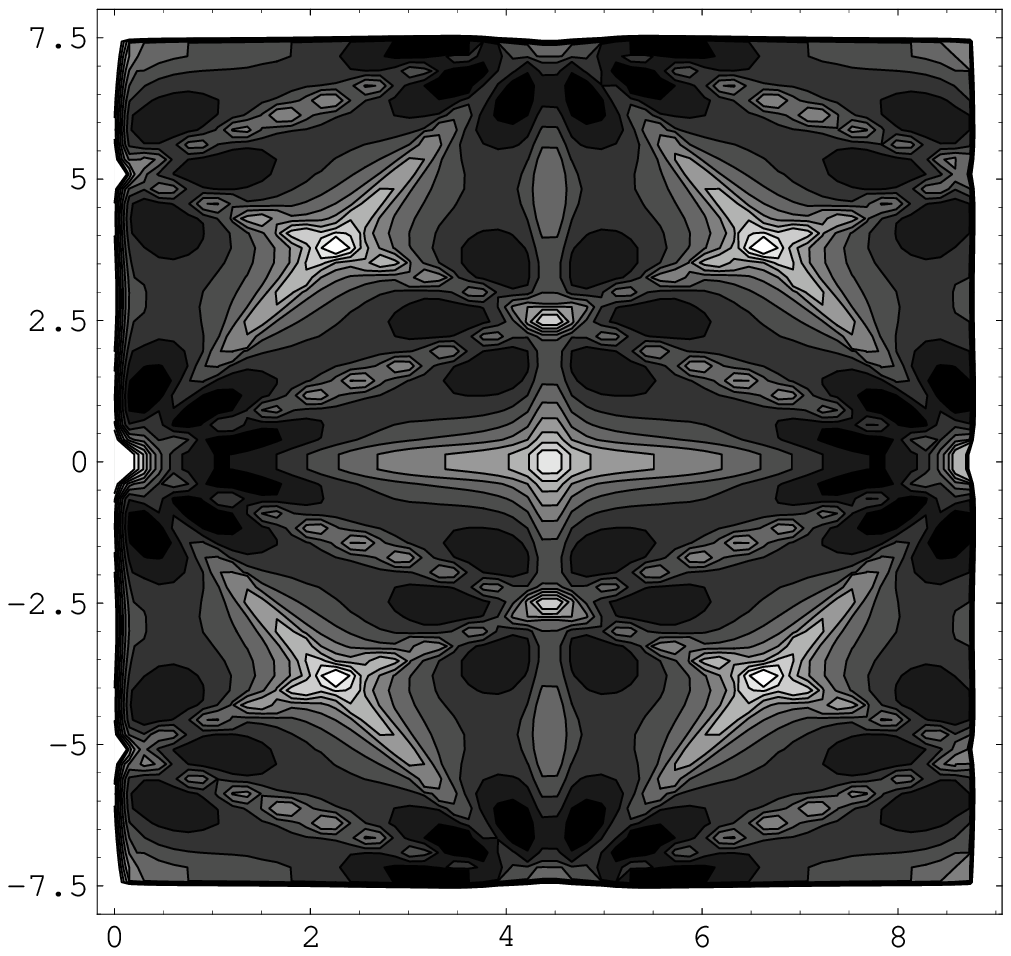} &\includegraphics[scale=0.45]{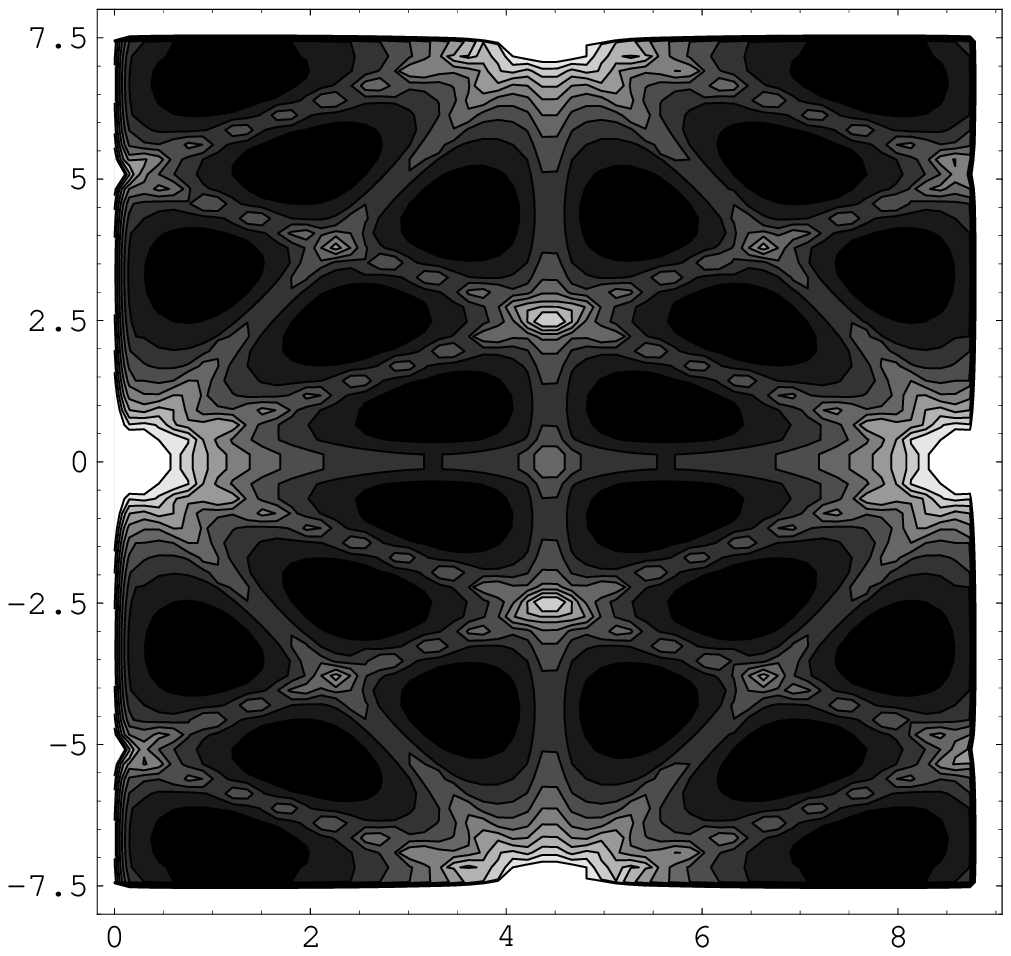} \\
(a) & (b) & (c)
\end{tabular}
\caption{\label{fig:g2potential} We show the $G_2$ one-loop potential on the Cartan torus for (a) high temperatures $\beta/R=0.01$, (b) intermediate temperatures $\beta/R=1.25$ and (c)low temperatures $\beta/R=100$. As we lower the temperature, the initial single minimum $A_0=0$ becomes a maximum and new non-trivial symmetrically located minima appear, being pushed apart. The appearance of strong repulsion lines along the symmetry axis is due to the logarithmic term. These two features do not appear in other topologies, like the torus \cite{Barbon:2006us, Pepe:2006er}.}
}

\section{Orthogonal and Symplectic Groups}
\label{son} 
In this section, we will consider the phase structure of $SO(2N)$ and $Sp(2N)$ gauge theories. The center of $Sp(2N)$ is $\bZ{2}$ and that of $SO(N)$ is $\bZ{2} \times \bZ{2}$ or $\bZ{4}$. for $N$ even and $\bZ{2}$ for $N$ odd.  However, since we are not considering spinors, only a $\bZ{2}$ subgroup of the center will be relevant for our analysis for orthogonal groups for even $N$. For odd $N$, $SO(N)$ groups have a trivial center.  

The non-trivial roots of $SO(2N)$ and $Sp(2N)$, in terms of the fundamental weights $\vec{\nu}$,  are given by $\pm \vec{\nu}_i \pm  \vec{\nu}_j$ (for $SO(2N)$, $i \neq j$ whereas there is no such restriction for $Sp(2N)$).  
For $SO(2N)$, using the gauge transformation, the Polyakov loop can be brought in the form:
\begin{equation}\label{sopolya}
	O = \begin{pmatrix}
		 \cos \theta_1  & \sin \theta_1 & & &  &&&\\
		-\sin \theta_1& \cos \theta_1 &  & &&&&\\
		&& \cos \theta_2  & \sin \theta_2  & &&&\\
		&&-\sin \theta_2 & \cos \theta_2  &&&&\\
		\phantom{\ddots} & \phantom{\ddots} & \phantom{\ddots}& \phantom{\ddots}& \ddots & \ddots && \\
		& & & & &&\cos \theta_N  & \sin \theta_N   \\
		& & & & && -\sin \theta_N  & \cos \theta_N    
		\end{pmatrix}\,,
\end{equation}
Also, by a gauge transformation we can flip the signs of $\theta$'s: $\theta_i \rightarrow -\theta_i$. 
The effective potential for the $\theta_i$'s  is given by
\begin{equation}\label{soeven}
V(\theta_i) = \sum_{i\neq j} {\cal V}(\theta_i - \theta_j) +{\cal V}(\theta_i + \theta_j).
\end{equation}

Introducing the density of eigenvalues $\rho(\theta)={1\over 2 \pi}+\sum_{n=1}^\infty \rho_n \cos n \theta $, (where the Fourier coefficient $\rho_n$ is now real and $\rho(\theta)=\rho(-\theta)$), and converting summations to integrals,  we obtain
\begin{equation}
V[\rho]=N^2\int_{-\pi}^\pi d \theta\rho(\theta) \int_{-\pi}^\pi d \theta' \rho(\theta') \sum_{n=1}^\infty {1\over n} (1-z_V(q^n)) (\cos(n(\theta-\theta'))+\cos(n(\theta+\theta'))).
\label{socos}
\end{equation}
Defining the Fourier components of $\rho(\theta)$ as in (\ref{fourier}), the potential can be written as
\begin{equation}
V[\rho_n]= N^2 \pi^2 \sum_{n=1}^\infty { (1-z_V(q^n)) \over n} \rho_n^2.
\label{soeffect}
\end{equation} 
For low temperatures, $\rho_n=0$, which implies a uniform distribution $\rho(\theta) = {1 \over 2 \pi}$ and a vanishing expectation value of the Polyakov loop. At extremely high temperatures, the arguments of the cosine  on the right hand side of  (\ref{socos}) have to be minimized, which happens when the $\theta_i'$ are the same and equal to 0 or $\pi$: {\em i.e.}
\[
\rho(\theta)= \delta(\theta)~~~ {\rm or} ~~~\delta ( \theta - \pi).
\]
In the two saddle points, the Polyakov loop $\tr O = 2N \int d \theta ~\rho(\theta) \cos(\theta)$ is
\begin{eqnarray*}
\tr O  & = & +1 ~~~~~~ \rho(\theta) = \delta ( \theta), \\
\tr O  & = & -1 ~~~~~~ \rho(\theta) = \delta ( \theta- \pi).
\end{eqnarray*}
If we sum over the different saddle points, we get a trivial value of the Polyakov loop. However, in the large N limit, we can choose one saddle point since large N provides a thermodynamic limit. In that case, the Polyakov loop is non-zero and the center symmetry is spontaneously broken. 

The situation is different for $SO(2N+1)$. The group $SO(2N+1)$ has no
center.  The roots, in terms of the fundamental weights $\vec{\nu}$, are
given by $ \pm \vec{\nu}_i \pm \vec{\nu}_j$ and $\pm \vec{\nu}_i$. 
The Polyakov loop is now a $(2N+1) \times (2N+1)$ matrix similar to
(\ref{sopolya}) with an extra row and column having 1 along the diagonal
and zeros everywhere else.
The effective potential for the $\theta_i$'s  is now given by
\begin{equation}
V(\theta_i) = \sum_{i \neq j} {\cal V}(\theta_i - \theta_j) +{\cal V}(\theta_i + \theta_j)+ 2\sum_i {\cal V}(\theta_i) \,. 
\end{equation}
As before, we can introduce a density of eigenvalues and its Fourier components yielding an effective potential (\ref{soeffect}).  The effective potential then is
\begin{eqnarray}
V[\rho]&=& N^2 \sum_{n=1}^\infty {1\over n} (1-z_V(q^n)) \Bigl( \int_{-\pi}^\pi d \theta\rho(\theta) \int_{-\pi}^\pi d \theta' \rho(\theta') \bigl(\cos(n(\theta-\theta'))+\cos(n(\theta+\theta')) \bigr)  \nonumber  \\ && ~~~~~~~~~~~~~~~~~~~+  { 1 \over N}  \int_{-\pi}^\pi d \theta \rho(\theta)  \cos(n\theta) \Bigr).
\label{soodd}
\end{eqnarray}
Defining the Fourier components of $\rho(\theta)$ as in (\ref{fourier}), the potential can be written as
\begin{equation}
V[\rho_n]= N^2 \pi^2 \sum_{n=1}^\infty { (1-z_V(q^n)) \over n} (\rho_n^2 +{ 1\over \pi N} \rho_n).
\label{sooddeffect}
\end{equation} 
This is minimized when $\rho_n = -{ 1 \over 2 \pi N}$. This corresponds to an eigenvalue distribution 
$\rho(\theta) = { 1\over 2\pi}(1 + { 1 \over 2N}) - { 1 \over 2N} \delta(\theta)$. The Polyakov loop is given by
\begin{equation}
\tr O =  1+2N \int_{-\pi}^\pi d \theta \rho(\theta)  \cos \theta = 1+2N \int_{-\pi}^\pi d \theta  \bigl( { 1\over 2\pi} (1  + {1 \over 2N})- { 1 \over 2N} \delta(\theta) \bigr) \cos \theta =  0.
\end{equation}

In the extreme high temperature phase, because of the second line in (\ref{soodd}) which picks out the $\theta=0$ saddle point, $\rho(\theta) = \delta(\theta)$, i.e. there is a unique saddle point in this case, giving a non-zero value of the Polyakov loop. The linear term on $\rho_n$ is similar to the one present for $SU(N)$ theories with fundamental matter. In that case, the potential becomes
\begin{equation}
V[\rho_n]= N^2 \pi^2 \sum_{n=1}^\infty { (1-z_V(q^n)) \over n} \rho_n^2 + {N_f\over N}{ z_F(q^n)\over n} \rho_n, \,
\end{equation} 
and is minimized at
\begin{equation}
\rho_n=-{N_f\over 2 N}{ z_F(q^n)\over 1-z_V(q^n)}\,.
\end{equation}

The value of the Fourier components is bounded to $|\rho_n|\leq 1$
because $\rho(\theta)\geq 0$. The bound is saturated at a temperature
below the Hagedorn temperature, and the eigenvalue density changes to a
gapped distribution in a continous way. It can be shown that the
deconfinement transition becomes Gross-Witten like
\cite{Schnitzer:2004qt, Skagerstam:1983gv}. However, in the case of
$SO(2N+1)$ theories, both the quadratic and the linear coefficient come
from the adjoint contribution and are the same. Hence $\rho_n=-1/(2\pi N)$ is
fixed and therefore the transition is Hagedorn-like. 

In the large $N$ limit, the behavior of the theory with gauge group
$Sp(2N)$ is identical the the $SO(2N)$ case.

\section{Gauss law and Screening}
\label{Gauss}
In the previous two sections, we find that  the Polyakov loop can be non-zero at high temperature is theories with no surviving center symmetry, even on finite volume.  We are introducing a non-Abelian charge in finite volume and it does not cost infinite free energy. This is consistent with the non-Abelian Gauss  law which we can write as
\begin{equation}
\int_{S^3}\left( \nabla_i E_i^a -f^{abc} A_i^b E_i^c \right) = Q^a,
\end{equation}
where $E_i^a$ are the electric components of the non-Abelian field, $A_i^a$ is the magnetic potential, $f^{abc}$ are the group structure constants and $Q^a$ is the total external charge. The first term on the left hand side  is a total derivative and vanish when integrated over a compact space.  This is the only term that is present for Abelian fields and implies the known result that we cannot introduce an Abelian charge on finite volume.  However, for non-Abelian fields the second term can be non-zero and Gauss law can be satisfied in the presence of a net external charge.

For a gauge theory with group $G$ and an unbroken center ${\cal C}$, both the measure and the action are invariant under center transformations. This implies that there is a {\em global} center symmetry which cannot be spontaneously broken in finite volume. Then, the expectation value of any center-breaking operator should be zero. The Polyakov loop in any representation charged under the center transforms non-trivially under the center (which can be seen by performing large gauge transformations that are periodic up to an element of the center). Therefore, its expectation value should be zero, which corresponds to the statement that the free energy cost of introducing external particles that are charged under the center is infinite. 
When there is no center symmetry because the gauge group is centerless or the representation of the matter fields explicitly breaks it, there is no symmetry protecting the expectation value of the Polyakov loop and indeed we observe that it is nonzero in general.

In the zero temperature limit, we find that the Polyakov loop is zero for any theory.  We can also understand this in terms of Gauss law. In this limit, we can ignore the Polyakov loop and  the term coming from the non-Abelian nature of the group in Gauss law. This is because we are essentially working in the zero coupling limit. The only non-trivial interaction terms we have kept correspond to the interaction of the fields with the background field $\alpha$ which is a zero mode of the theory. \footnote{In a normalization of fields that correspond to canonical kinetic terms in the action, this corresponds to taking the $g \rightarrow 0$ limit after rescaling $\alpha \rightarrow {1 \over g} \alpha$. } There is no $\alpha$ at  zero temperature and therefore no non-Abelian term in Gauss law.  In this case, it costs infinite action to introduce any external charge and the Polyakov loop expectation value is zero. 

This is consistent with the physical picture in terms of screening. In the absence of a surviving center symmetry, an external particle in any representation can be screened by an appropriate number of gluon or matter particles, which allows us to introduce them at finite free energy cost. In theories with a center symmetry, an external particle charged under the center cannot be screened and therefore cannot have finite free energy in finite volume. 

\section{Conclusions}
\label{conclusions}

In this paper, we have investigated confinement on  $S^3 \times S^1$ for a gauge theory based on an arbitrary semi-simple Lie group containing matter in generic representations.  Our analytic calculation reproduces the behavior of the Polyakov loop computed by lattice simulations for any theory for which lattice data exist.\cite{Greensite:2006sm,Pepe:2006er,Cossu:2007dk,Barresi:2003jq,Holland:2003kg}.

By using the general expression for the one-loop effective potential, we have shown that at zero temperature the Polyakov loop is exactly zero for any theory.  Since the Polyakov loop is naturally related to the free energy of an isolated quark, our result implies that on a $S^3 \times S^1$ manifold, any gauge theory confines. Although in this case confinement is due to kinematics, this finding is likely to be the weak coupling equivalent of confinement in gauge theories. For small but non-zero temperatures, the Polyakov loop is still zero if the theory has an unbroken center symmetry, but acquire a non-zero (albeit exponentially small ${\cal O}(e^{-\beta /R})) $ vacuum expectation value otherwise. The very same features characterise the high temperature phase, although their explicit realisation differ from the low temperature regime. In the case in which the system is center symmetric, this symmetry cannot be broken on a finite volume, hence the Polyakov loop will still be zero at high temperature. However, this zero is due to a different phenomenon, namely the emergence of center-breaking minima of the effective potential, over which we have to take an average when we are not the in the thermodynamic limit. In this limit one minimum is selected, and this gives a non-zero value to the Polyakov loop. This is the case for $SU(N)$ and $SO(2N)$ gauge theories. For centerless theories, like $G_2$ and $SO(2N+1)$, there seems to be only one minimum for the effective potential in the high temperature phase. On this saddle point, the Polyakov loop is different from zero.

In this paper, we have not considered Polyakov loop in the spinor representations of $SO(N)$. Since these Polyakov loops will be non-trivially charged under the center of the gauge group, they should exhibit a behavior similar to the fundamental Polyakov loop in the $SU(N)$ theory, i.e. at finite $N$, they should be zero at all temperatures. In the thermodynamic (large $N$) limit, after restriction to one center-breaking minimum, we should find a non-zero value of the Polyakov loop. We have also not studied the Polyakov loop in the adjoint representation of $SU(N)$ theories. We expect to find that all representations uncharged under the center would have a non-zero Polyakov loop at non-zero temperature with a smooth cross over between the low temperature and the high temperature regimes. 

The minima of the effective potential exhibit a different structure in the low and the high temperature phase. This is true independently of the existence of an unbroken center group of the theory. This means that with a finite number of degrees of freedom, a cross over between the confined phase and the deconfined phase should take place. In the large $N$ limit, this cross-over becomes a real phase transition. For finite $N$ or in the case of exceptional groups, whether there is a cross-over or a real phase transition cannot be established by a computation on  $S^3 \times S^1$. Recent lattice calculations suggest that in pure gauge theories the change of properties of the vacuum is always associated with a proper phase transition \cite{Pepe:2006er,Cossu:2007dk,Barresi:2003jq,Holland:2003kg}.

Another interesting question is whether calculations on a $S^3 \times S^1$ manifold can shed some light over the mechanism of color confinement. For QCD with matter in the adjoint representation, an interesting development in this direction is reported in \cite{Unsal:2007jx}.

\acknowledgments
We would like to thank Guido Cossu and Prem Kumar for valuable discussions.  BL is supported by a Royal Society University Research Fellowship and  AN is
supported by a PPARC Advanced Fellowship. 
\bibliographystyle{myutcaps}
\bibliography {s3cs1} %
\end{document}